\documentclass[journal=ancac3,manuscript=article,layout=twocolumn]{achemso}
\usepackage[T1]{fontenc}
\usepackage{graphicx} 
\usepackage{siunitx} %
\usepackage{xspace} 
\usepackage{float}
\usepackage{hyperref}
\usepackage{pdfpages}

\newcommand{\fz}{\ensuremath{f_0}\xspace}
\newcommand{\Bz}{\ensuremath{B_0}\xspace}
\newcommand{\be}{\ensuremath{b}\xspace}
\newcommand{\w}{\ensuremath{w}\xspace}

\title{Efficient Generation of Second-Harmonic Propagating Spin Waves in a Thin, Out-of-Plane-Magnetized Ferromagnetic Film}


\author{Mathieu Moalic}
\email{matmoa@amu.edu.pl}
\affiliation{Institute of Spintronics and Quantum Information, Faculty of Physics and Astronomy, Adam Mickiewicz University, 61-614 Poznań, Poland}

\author{Youenn Patat}
\affiliation{Institute of Spintronics and Quantum Information, Faculty of Physics and Astronomy, Adam Mickiewicz University, 61-614 Poznań, Poland}

\author{Mateusz Zelent}
\affiliation{Institute of Spintronics and Quantum Information, Faculty of Physics and Astronomy, Adam Mickiewicz University, 61-614 Poznań, Poland}

\author{Maciej Krawczyk}
\affiliation{Institute of Spintronics and Quantum Information, Faculty of Physics and Astronomy, Adam Mickiewicz University, 61-614 Poznań, Poland}

\begin{document}

\begin{abstract}
Spin waves are attractive information carriers owing to their gigahertz-to-terahertz frequencies, nanometric wavelengths, and negligible Joule heating. Yet the efficient excitation of short-wavelength, high-frequency spin waves and the exploitation of nonlinear effects remain challenging. We propose a hybrid ferromagnetic nanostructure composed of a small, in-plane-magnetized rim (a magnonic nanocavity) exchange-coupled to an out-of-plane-magnetized region. Micromagnetic simulations show that a spatially uniform out-of-plane microwave field excites the rim’s fundamental mode; its second harmonic is then coherently and efficiently launched into the second region of the structure, yielding propagating spin waves. The process can be realized in strip or disk geometries, providing excitation of plane-wave or radial spin waves, respectively. The conversion efficiency grows nonlinearly with the pump amplitude and can be further improved when the frequency of a higher-order standing wave in the nanocavity matches the second-harmonic frequency. The emission frequency is tunable via the bias magnetic field or the width of the nanocavity, suggesting a compact route toward on-chip, short-wavelength, high-frequency spin-wave sources for artificial neural networks.
\end{abstract}

\maketitle


Although the first demonstration of the doubling of microwave frequency in ferrites dates back to the 1950s \cite{Ayres1956}, second-harmonic generation (SHG) and, more generally, multi-harmonic generation of spin waves (SWs) have advanced rapidly in the past decade. 
This progress is mainly due to their potential applications in magnonic neuromorphic computing \cite{Papp2021,Korber2023,Fripp2023,Marrows2024}, because, as with any neural network, a nonlinear response to the pump amplitude is a crucial property.

It has been shown that edge SWs, which are confined by an inhomogeneous demagnetizing field \cite{Sebastian2013}, and the oscillations of a pinned domain wall \cite{Hermsdoerfer2009}, can act as nonlinear emitters that radiate SWs at twice the drive frequency. 
Domain walls and related magnetic textures (e.g., vortices and skyrmions), besides serving as guides for short-wavelength magnons \cite{Wagner2016,Rodrogues2021}, act as natural loci of broken symmetry that enhance frequency multiplication \cite{Sluka2019}.
However, in most cases the SWs generated in the SHG process remain confined to the magnetic textures \cite{Rodrogues2021}.
More recently, dispersion- and symmetry-engineered nanowaveguides have enabled phase-matched, resonant SHG of propagating magnons, establishing efficient conversion into the $2f$ branch \cite{Nikolaev2024,Nikolaev2025APL}. The anisotropic dispersion relation of magnetostatic SWs \cite{stancilprabhakar} allows for SHG of SWs with positive \cite{Nikolaev2024} or negative \cite{Nikolaev2025APL} group velocities in the Damon–Eshbach configuration (magnetization perpendicular to the wave vector) and the backward-volume configuration (magnetization parallel to the wave vector), respectively. 
 
In parallel, hybrid quantum cavity–magnonic systems have shown that magnon nonlinearities support coherent harmonic generation and broadband microwave frequency conversion—with phase-preserving harmonics and multiwave mixing—highlighting opportunities for on-chip signal processing and quantum control \cite{Wu2024,Lan2025}. 

Together, these results position SHG as a practical route to generating coherent, high-frequency, short-wavelength magnons from uniform microwave drives, provided that dispersion (phase matching) and mode symmetry are tailored appropriately \cite{Nikolaev2024,Nikolaev2025APL}.

There are also many other methods for generating short SWs (i.e., sub-\SI{100}{\nano\meter}) using linear processes, which are desirable for analog and binary magnonic devices \cite{Chumak2022rev}. Existing strategies include (a) microwave nanostrip antennas \cite{Ciubotaru2016}, (b) magnetic solitons such as domain walls and vortices \cite{Woo2017,Wintz2016,Dieterle2019}, (c) magnetostatic coupling from ferromagnetic stripes or nanocontacts \cite{Liu2018}, (d) two-dimensional grating couplers \cite{Yu2016}, and (e) internal-field engineering via lithographic modulation of demagnetizing fields \cite{Mushenok2017}. Most of these approaches face trade-offs: they either operate at relatively low frequencies, require large DC currents, or impose stringent lithographic constraints. 
Even the recent demagnetizing-field–modulation technique in fully perpendicular films has reached wavelengths down to \SI{200}{\nano\meter} only at modest frequencies and in carefully tailored geometries \cite{Wang2023}.
Harnessing nonlinear harmonic conversion may circumvent these limitations \cite{Kumar2024}, especially when implemented in a lithographically simple geometry.


\section{Results}
\begin{figure}
    \centering
    \includegraphics[width=0.99\linewidth]{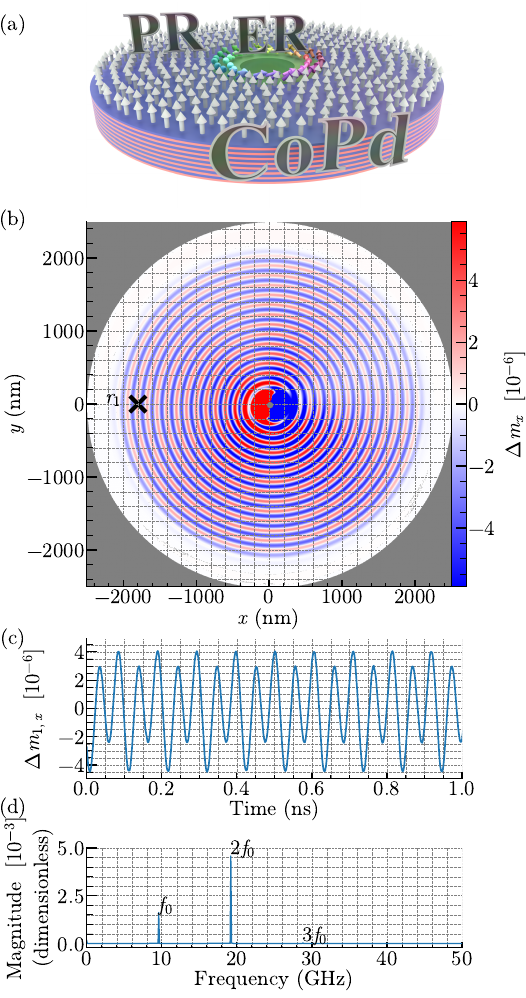}
    \caption{
        (a) Schematic illustration of the investigated structure showing a Co/Pd disk. Note that the figure is not to scale. The green area around the hole represents the ER with reduced PMA. The arrows indicate the orientation of the magnetization.
        (b) Snapshot of the two-dimensional simulation at steady state for an excitation frequency of $f = \SI{8.60}{\giga\hertz}$. 
        (c) Time evolution of the magnetization at point $r_1$, indicated in panel (b).
        (d) Discrete Fourier transform of the magnetization signal at $r_1$ over time.
    }
    \label{fig:1}
\end{figure}

\begin{figure}
    \centering
    \includegraphics[width=0.99\linewidth]{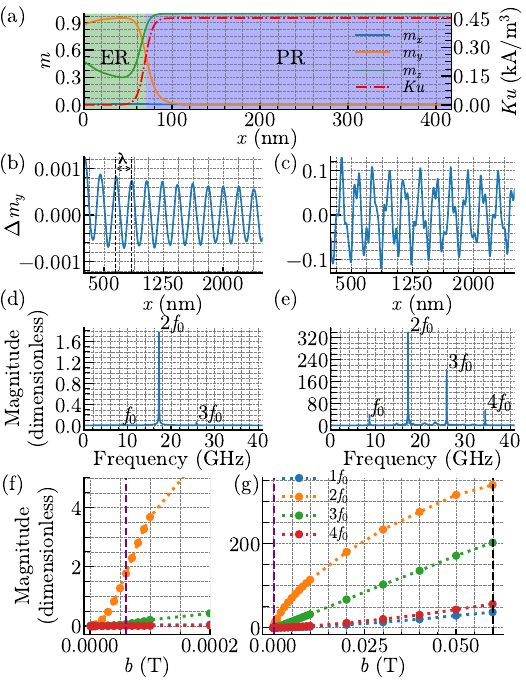}
    \caption{
        (a) Static magnetization components and out-of-plane anisotropy value along the $x$ axis (for $x < \SI{400}{\nano\meter}$).
        (b) Dynamical $y$-component of the magnetization along the waveguide under a uniform out-of-plane drive at $\fz=\SI{8.60}{\giga\hertz}$ with microwave magnetic field amplitude $b_1=\SI{60}{\micro\tesla}$. The wavelength $\lambda$ corresponds to the $2\fz$ component.
        (c) Same as (b), but with an amplitude of $b_2 = \SI{60}{\milli\tesla}$.
        (d) Discrete Fourier transform of the $y$-component of the magnetization in a cell located \SI{2}{\micro\meter} from the excitation source, for excitation amplitude $b_1$.
        (e) Same as (d), but for $b_2$.
        (f,g) Magnitudes of the Fourier transform peaks at 2\fz, 3\fz, and 4\fz as a function of $\be$. The values $b_1$ and $b_2$, used in panels (b,d) and (c,e) respectively, are marked with dashed lines.
    }
    \label{fig:2}
\end{figure}

We unveil a strategy for locally exciting coherent, high-frequency, propagating SWs through SHG, driven by a simple, uniform, out-of-plane microwave field. The key idea lies in a hybrid nanostructure engineered to have two distinct magnetic regions: an ``Excitation Region'' (ER) with in-plane magnetization and a ``Propagation Region'' (PR) with perpendicular magnetic anisotropy (PMA). These regions are connected by a \ang{90} domain wall. We demonstrate that the effect exists in a 2D geometry—yielding radially propagating SWs from a centrally localized, in-plane-magnetized rim (Fig.~\ref{fig:1})—and in 1D, resulting in the emission of plane waves (Fig.~\ref{fig:2}). 
A detailed analysis reveals that the in-plane–magnetized ER acts as a nonlinear nanocavity that up-converts the uniform drive to 2\fz, while the ER–PR interface efficiently couples out the resulting wave into the PR. SHG is particularly effective when the width and bias magnetic field are tuned to match the formation of standing SWs in the nanocavity at 2\fz. This provides a direct route for the local generation of technologically vital, high-frequency, short-wavelength magnons, especially for magnonic artificial neural networks and quantum magnonics.

\paragraph{Disk Geometry.}
For this study, we selected a [Co/Pd]$_8$ multilayered film of \SI{13.2}{\nano\meter} thickness in which we introduced two distinct regions differing only in their PMA. The first region, ER, lacks uniaxial anisotropy and exhibits in-plane magnetization at remanence (see the green area in Fig.~\ref{fig:1}(a) and Fig.~\ref{fig:2}(a)). The second region, PR, exhibits sufficiently strong PMA to saturate the magnetization out of plane (see the blue area in Fig.~\ref{fig:1}(a) and Fig.~\ref{fig:2}(a)). The PMA profile is engineered to transition smoothly from zero to the bulk value over a \SI{50}{\nano\meter} region; this profile is shown in Fig.~\ref{fig:2}(a) as the red dashed line (the same profile is applied radially in the 2D system).

A bias magnetic field \Bz was applied perpendicular to the film plane (along the $z$-axis) up to \SI{0.6}{\tesla}. Within this field range, the magnetization in the ER gradually cants away from the in-plane orientation, changing from \ang{7.1} at $\Bz = \SI{200}{\milli\tesla}$ to \ang{36} at $\Bz = \SI{600}{\milli\tesla}$. The two regions are connected by an approximately \ang{90} domain wall at remanence. The material parameters and the effective model used to simulate the multilayer are described in the Methods section.

To study SW dynamics, we applied a microwave magnetic field (\be) oriented out of plane (along the $z$-axis). This induces a torque that acts only on the magnetization in the ER and the domain wall while leaving the PR unexcited.
Simulations were conducted using Amumax~\cite{amumax2023}, our fork of MuMax3, to solve the Landau–Lifshitz–Gilbert equation. Details of the simulations are provided in the Methods section.

We started the simulations with the film shaped as a thin disk (radius \SI{2.5}{\micro\meter}) containing a central antidot (hole) with a diameter of \SI{80}{\nano\meter}, surrounded by a rim (an in-plane-magnetized ER of width $\w = \SI{40}{\nano\meter}$). A high-damping region with a width of \SI{500}{\nano\meter} was included near the outer edge of the disk to suppress SW reflections.

Under an external magnetic field of $\Bz = \SI{354}{\milli\tesla}$, the lowest resonant mode in the PR appeared at \SI{13.90}{\giga\hertz}. However, lower-frequency modes still existed within the ER. A sinc-type microwave field was first applied along the $z$-axis with cutoff frequency $f_\text{cut} = \SI{10}{\giga\hertz}$ and a peak magnitude of \SI{10}{\milli\tesla} to excite SWs. The only peak below $f_\text{cut}$ was found at $\fz = \SI{8.60}{\giga\hertz}$. 
We then ran a steady-state simulation with a sinusoidal excitation at \fz and a microwave field amplitude of $\be = \SI{100}{\micro\tesla}$. Figure~\ref{fig:1}(b) presents a snapshot of the magnetization deviation from the ground state, $\Delta m_x$, over the whole sample, revealing radially propagating SWs, i.e., a pattern similar to that observed in the disk in the vortex state, excited by vortex gyration \cite{Dieterle2019,Wintz2016}. 

To analyze the spectral content, we monitored the dynamical component of the reduced magnetization along $x$, $\Delta m_x$ (defined as the deviation from the ground-state magnetization), over time at a point $r_1$ located \SI{1800}{\nano\meter} from the disk center. The resulting time trace is shown in Fig.~\ref{fig:1}(c), and its discrete Fourier transform is presented in Fig.~\ref{fig:1}(d).
This confirms that the dominant radiation frequency is 2\fz (magnitude of $4.6 \times 10^{-3}$), with much weaker components at \fz ($1.6 \times 10^{-3}$) and 3\fz ($0.1 \times 10^{-3}$), respectively. Because the excitation frequency lay below the ferromagnetic resonance (FMR) of the PR, the signal at \fz is an evanescent wave; the strong 2\fz signal is therefore notable given the moderate excitation amplitude. This behavior contrasts with SW emission from the vortex core at the pumped frequency \cite{Dieterle2019,Wintz2016}.


\paragraph{One-Dimensional Waveguide.}
To explain the mechanisms behind higher-order SW generation, we simplified the system by designing a one-dimensional geometry of length \SI{3600}{\nano\meter} along the $x$-axis, with a single simulation cell along $y$ and periodic boundary conditions implemented as 1{,}000 replicas on each side. The width of the ER was $\w=\SI{70}{\nano\meter}$, which was sufficient to ensure that the magnetization remained mostly in plane and aligned along the $+y$ axis (Fig.~\ref{fig:2}(a)).

From simulations at a static magnetic field $\Bz=\SI{354}{\milli\tesla}$ with the sinc-type excitation and the out-of-plane-oriented microwave magnetic field \be, we found that the strongest response in the ER, which dominated the spectrum, occurred at $\fz=\SI{8.60}{\giga\hertz}$ (the spectrum is shown in Fig.~S1 of the Supporting Information, SI).
\paragraph{Nonlinear Regime and Higher Harmonics.}
We then performed steady-state simulations using a sinusoidal microwave field of frequency $\fz=\SI{8.60}{\giga\hertz}$ and amplitude $\be=b_1=\SI{60}{\micro\tesla}$. Although this excitation frequency lay below the FMR of the PR (i.e., \SI{13.90}{\giga\hertz}), a strong 2\fz SW signal was observed propagating in the PR along the $x$-axis with wavelength $\lambda = \SI{261}{\nano\meter}$, as shown in Fig.~\ref{fig:2}(b). The discrete Fourier transform of the magnetization at $x = \SI{2}{\micro\meter}$ (Fig.~\ref{fig:2}(d)) revealed a pronounced peak at 2\fz and very weak contributions at \fz and 3\fz, in agreement with the 2D system.

To further probe the nonlinear-dynamics regime, we repeated the simulation with a stronger excitation field, $\be = b_2 = \SI{60}{\milli\tesla}$ (an increase by a factor of 1{,}000), as shown in Fig.~\ref{fig:2}(c). In this case, the system generated a more complex signal that propagated in the PR. In addition to the dominant peak at 2\fz (magnitude 340 arb. units), higher-frequency harmonics appeared in the spectrum at 3\fz (200 arb. units) and 4\fz (60 arb. units), as shown in Fig.~\ref{fig:2}(e).

Figure~\ref{fig:2}(f,g) shows how the amplitudes of the harmonics generated in the PR evolved as the excitation amplitude increased from \SI{1}{\micro\tesla} to \SI{60}{\milli\tesla} (2\fz, 3\fz, and 4\fz). The second-harmonic amplitude increased quadratically with $\be$ up to $\be=\SI{0.1}{\milli\tesla}$, indicating a thresholdless process, similar to other SHG scenarios \cite{Demidov2011,Cheng2013}.
For larger microwave amplitudes, the amplitude of the 2\fz wave scaled as $\sqrt{\be}$ and saturated at high $\be$. This behavior indicates the onset of additional threshold processes, which became relevant when the magnetic field exceeded \SI{0.1}{\milli\tesla}, as evidenced by the multiple low-intensity peaks visible in Fig.~\ref{fig:2}(e), generated in higher-order nonlinear processes \cite{Schultheiss2012,Lan2025}. At these excitation levels, the amplitudes of the third and fourth harmonics increased significantly, narrowing the gap relative to the 2\fz wave. This demonstrates the potential for enhancing the amplitude of propagating SWs at twice the frequency of the pumping microwave field, although the saturation mechanism becomes significant at larger excitation amplitudes.

We also studied the effect of damping (see Fig.~S2 in the SI) and found that increasing the damping from $\alpha=\num{1e-3}$ to $\alpha=\num{1e-2}$ fully suppressed higher-harmonic generation. Raising the driving field from $\SI{6}{\micro\tesla}$ to $\SI{60}{\micro\tesla}$ restored a clean 2\fz response.
In metallic PMA multilayers (e.g., Pt/Co, Co/Ni, Co/Pd, CoFeB/Au), room-temperature Gilbert damping is typically of order $10^{-2}$ and often $\alpha \approx 0.01\text{–}0.05$; it is due to interfacial spin–orbit and spin-pumping contributions \cite{Kato2012,Song2013,Barati2017,Kuswik_2017,Janardhanan2023}. By contrast, YIG films routinely show $\alpha \sim 10^{-4}$ and can approach a few $\times 10^{-5}$ in thicker or optimized samples \cite{Sun2012,Hauser2016,Soumah2018}. Thus our baseline $\alpha=10^{-3}$ is optimistic for Co/Pd by one order of magnitude but higher than low-loss garnets. This means that, experimentally, similar $2\fz$ emission in metallic stacks, as presented above, will generally require larger drive fields and/or materials optimization to reduce $\alpha$.


\begin{figure}
    \centering
    \includegraphics[width=0.99\linewidth]{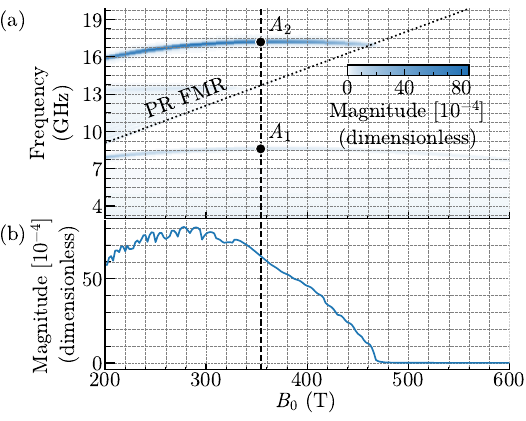}
    \caption{
        (a) Spin-wave spectrum in the propagation region (PR) recorded at $x=\SI{2000}{\nano\meter}$ while sweeping the bias field \Bz. For each \Bz, the system is driven by a uniform out-of-plane sinusoidal field at the ER fundamental frequency $\fz(\Bz)$ with fixed amplitude $\be=\SI{100}{\micro\tesla}$; the dotted line marks the PR FMR. The vertical dashed line indicates the operating field used elsewhere in the paper ($\Bz=\SI{354}{\milli\tesla}$). Markers $A_1$ and $A_2$ denote SW modes that are analyzed in Fig.~\ref{fig:4}.
        (b) Magnitude of the 2\fz Fourier component at $x=\SI{2000}{\nano\meter}$ versus \Bz under the same uniform out-of-plane sinusoidal drive at $\fz(\Bz)$ with $\be=\SI{100}{\micro\tesla}$.
    }
    \label{fig:3}
\end{figure}
We then explored two strategies to control the SW generation frequency \fz and its second harmonic: adjusting the static external field \Bz and varying the ER width \w.

\paragraph{Bias-Field Tuning.}
Figure~\ref{fig:3}(a) shows the PR spectrum at $x=\SI{2000}{\nano\meter}$ as \Bz was ramped from \SIrange{200}{600}{\milli\tesla}. For each \Bz, we recomputed the ER fundamental $\fz(\Bz)$ and drove the system with a uniform out-of-plane sinusoid at this frequency with fixed amplitude $\be=\SI{100}{\micro\tesla}$—that is, only the pump frequency changed with \Bz; \be was kept constant. Within this field interval, the domain-wall center, defined as the point where $m_z=0.5$ (i.e., halfway between in-plane and out-of-plane), shifted from $x=\SI{58}{\nano\meter}$ to $x=\SI{64}{\nano\meter}$.
We recorded the PR response at $x=\SI{2000}{\nano\meter}$, i.e., far away from the ER. 
The PR FMR (dotted line) stays above \fz throughout. The 2\fz branch tracks \fz, increasing from \SIrange{16.0}{17.25}{\giga\hertz} as \Bz rises from \SIrange{200}{354}{\milli\tesla}, then bending down and crossing the PR FMR at \Bz=\SI{464}{\milli\tesla} ($2\fz=\SI{16.5}{\giga\hertz}$). For higher \Bz, 2\fz falls below the PR cutoff and the corresponding wave in the PR becomes evanescent.
One interesting feature is how the magnitude of the 2\fz branch changes with increasing \Bz. As shown in Fig.~\ref{fig:3}(b), it increases from 60 [arb. units] to a maximum of 81 [arb. units] at $\Bz=\SI{354}{\milli\tesla}$. As \Bz increases further and approaches the FMR frequency of the PR at $\Bz=\SI{464}{\milli\tesla}$, the SW amplitude decreases. At this frequency, the SW in the PR has an infinite wavelength, which causes its excitation efficiency to drop to zero.
This approach offers only moderate 2\fz tunability: a \SI{1.25}{\giga\hertz} range for a \SI{160}{\milli\tesla} change in the bias field, accompanied by concurrent moderate changes in the emitted intensity.


\begin{figure}
    \centering
    \includegraphics[width=0.99\linewidth]{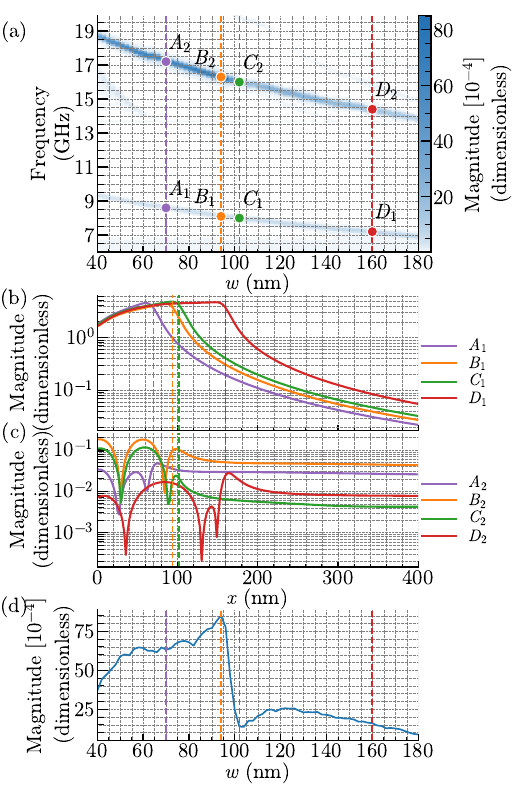}
    \caption{
        (a) Evolution of the SW spectrum in the PR as a function of the ER width \w. 
        (b) Spatial profiles of the normalized SW amplitude corresponding to the \fz modes at different values of \w as marked in panel (a). The dashed lines represent the corresponding ER length \w.
        (c) Same as (b) but for the 2\fz frequency.
        (d) Magnitude of the 2\fz peak in the PR, for $x=\SI{2000}{\nano\meter}$ as \w is increased.
        The results are obtained with bias magnetic field $\Bz=\SI{354}{\milli\tesla}$, excited with the sine microwave magnetic field of magnitude $\be = \SI{100}{\micro\tesla}$.
    }
    \label{fig:4}
\end{figure}
\paragraph{Cavity-Width Tuning.}
A complementary tuning route is shown in Fig.~\ref{fig:4}(a): we fixed \Bz = \SI{354}{\milli\tesla} and varied the ER width, \w, from \SIrange{40}{180}{\nano\meter}. In these simulations, for each value of \w we recomputed the ER fundamental resonance \fz, drove the system with a sinusoid at that frequency with fixed amplitude $\be=\SI{100}{\micro\tesla}$, and recorded the PR response at $x=\SI{2000}{\nano\meter}$. As \w increases from \SIrange{70}{160}{\nano\meter}, \fz decreases from \SIrange{8.60}{7.44}{\giga\hertz} (and thus 2\fz from \SIrange{17.20}{14.88}{\giga\hertz}). 

In Fig.~\ref{fig:4}(d), we show the magnitude of the 2\fz branch as a function of \w. The magnitude first grows with \w—rising from $36\times10^{-4}$ at $\w=\SI{40}{\nano\meter}$ to $85\times10^{-4}$ at $\w=\SI{94}{\nano\meter}$—then drops to $14\times10^{-4}$ at $\w=\SI{102}{\nano\meter}$ and thereafter increases but remains below $26\times10^{-4}$. Unlike Fig.~\ref{fig:3}(b), the FMR frequency of the PR is essentially independent of \w, so 2\fz remains above the PR FMR across the entire range considered.
This non-monotonicity of the second-harmonic intensity therefore points to a resonance in the ER, with a maximum near $\w=\SI{94}{\nano\meter}$ (i.e., point $B_2$). 

\paragraph{Spatial distribution of SW modes.}
To clarify the origin of the \fz and 2\fz modes, we plot in Fig.~\ref{fig:4}(b) and (c) log-scaled maps of the SW-amplitude spatial distribution for the \fz and 2\fz branches, respectively, at the ER widths marked in Fig.~\ref{fig:4}(a). The fundamental \fz modes—$A_1$ ($f=\SI{8.60}{\giga\hertz}$, $\w=\SI{70}{\nano\meter}$), $B_1$ ($f=\SI{8.08}{\giga\hertz}$, $\w=\SI{94}{\nano\meter}$), $C_1$ ($f=\SI{8.00}{\giga\hertz}$, $\w=\SI{102}{\nano\meter}$), and $D_1$ ($f=\SI{7.20}{\giga\hertz}$, $\w=\SI{160}{\nano\meter}$)—exhibit essentially the same spatial pattern for all $\w$ (Fig.~\ref{fig:4}(b)): the amplitude is tightly confined to the ER, resembling a quarter wavelength, peaks at the domain wall, and decays monotonically into the PR, as expected for \fz lying below the FMR frequency of the PR. This invariance and below-cutoff character rule out the directly pumped \fz mode as the source of the observed resonance at $\w=\SI{94}{\nano\meter}$. The resonant response must therefore be associated with the 2\fz mode.

To support this hypothesis, we show the 2\fz mode profiles in Fig.~\ref{fig:4}(c): $A_2$ ($f=\SI{17.2}{\giga\hertz}$, $\w=\SI{70}{\nano\meter}$), $B_2$ ($f=\SI{16.16}{\giga\hertz}$, $\w=\SI{94}{\nano\meter}$), $C_2$ ($f=\SI{16.00}{\giga\hertz}$, $\w=\SI{102}{\nano\meter}$), and $D_2$ ($f=\SI{14.40}{\giga\hertz}$, $\w=\SI{160}{\nano\meter}$).
Unlike the \fz modes, the 2\fz modes have nodal points in the ER. Clearly, the SW magnitude is highest at $\w=\SI{94}{\nano\meter}$, in both the ER and the PR. Moreover, this is the only case in which the maximal amplitudes are formed at both edges of the ER. This confirms the formation of a standing wave at 2\fz and a resonant condition at $\w=\SI{94}{\nano\meter}$. This conclusion is further supported by tracking the evolution of this standing mode with changes in \w, as shown in Fig.~3 of the SI. We attribute the asymmetric profile of the standing wave to the continuous variation in the static magnetization orientation in the ER.

Another interesting aspect of the dependence shown in Fig.~\ref{fig:4}(c) is the drastic decrease in the 2\fz-mode magnitude at $\w=\SI{102}{\nano\meter}$, i.e., just after resonant excitation. Such a dependence resembles a Fano resonance—i.e., a resonant interaction between a confined mode and a continuous band—widely explored in photonics \cite{Limonov2017}. However, in our case it would arise from the interaction of the bound SW generated in the SHG process in the ER with the continuous band of propagating SWs in the PR.

To further ensure that the higher-frequency generation is due to a cavity effect in the ER rather than domain-wall excitation, as in other studies \cite{Hermsdoerfer2009,Wagner2016,Rodrogues2021,Sluka2019}, we ran a simulation with frozen spins (after relaxation) in the left part of the ER ($x<\SI{40}{\nano\meter}$) for $\w=\SI{70}{\nano\meter}$ (see Fig.~S1 in the SI). The spectral plot shows no higher-harmonic generation, allowing us to conclude that the nanocavity formed by the ER plays a crucial role between the ferromagnetic edge and the domain wall.

\section{Conclusions}

These results allow us to conclude that the discovered process exists in both 1D and 2D and enables the excitation of propagating SWs in the form of radial or plane waves, respectively. The fundamental (pumped) mode is spatially trapped in the ER and peaks at the ER–PR interface, decaying evanescently into the PR. By contrast, the second harmonic is a propagating mode in the PR that carries energy across the entire waveguide. Crucially, the ER acts as a resonant nanocavity, accumulating energy from the uniform pump at the fundamental mode frequency \fz, which up-converts the pumped energy nonlinearly to 2\fz (or higher multiples of \fz) and efficiently launches it into the adjacent PR. This process becomes especially efficient when the 2\fz frequency matches that of a higher-order standing wave of the nanocavity.

In summary, we have demonstrated, through comprehensive micromagnetic simulations, a compact scheme for generating coherent, propagating SWs at the second harmonic of the pumping field in a hybrid structure composed of two orthogonally magnetized parts: a nanocavity and an extended part. The key ingredients—a uniform out-of-plane microwave pump, an in-plane-magnetized ER, and an adjacent out-of-plane-magnetized PR—cooperate to (i) confine the fundamental pumping energy in the ER, (ii) up-convert its energy to 2\fz inside a nanocavity region, and (iii) launch the resulting short-wavelength wave into the PR. Crucially, the emission frequency can be tuned continuously either by sweeping the bias field \Bz or by adjusting the ER width \w, while the conversion efficiency increases nonlinearly with the pump amplitude. The concept extends naturally from one- to two-dimensional geometries, where radially propagating second-harmonic waves emerge from an in-plane-magnetized rim of the central antidot.

The proposed hybrid structure can be fabricated using any ferromagnetic thin film with PMA, e.g., metallic multilayers \cite{Sbiaa2011} or garnet thin films \cite{Das2024}, with an in-plane-magnetized nanocavity formed using one of the existing methods that allow the reduction of PMA locally. Examples of these methods include ion irradiation \cite{Wawro2018,Das2024}, oxidation \cite{Luo2019}, direct-write laser
exposure annealing \cite{Ridd2024}, or even local electric-field modulation \cite{Rana2019}. The final method suggests that the nanocavity may be tunable not only by an external magnetic bias field, as shown in our study, but also locally and in a more integrated way.
Thus, our results point to a lithographically simple, energy-efficient pathway toward on-chip sources of nonlinearly generated SWs with sub-\SI{300}{\nano\meter} wavelengths—an essential building block for next-generation magnonic logic, signal processing, and hybrid magnonic–quantum platforms. 
Future work will focus on optimizing material stacks to reduce damping (e.g., doped-yttrium iron garnet films), investigating the role of domain-wall chirality, and integrating the proposed emitter with magnonic networks and cavity–magnonic devices.


\section{Methods}
\paragraph{Material Parameters.}
We modeled a multilayered [Co/Pd]$_8$ system as a \SI{13.2}{\nano\meter}-thick effective Co layer with parameters taken from the literature~\cite{lemesh2018twisted,Pan2020c}: $K_{\text{u}} = \SI{4.5e5}{\joule\per\meter\cubed}$, $M_{\text{s}} = \SI{0.81e6}{\ampere\per\meter}$, $A_{\text{ex}} = \SI{1.3e-11}{\joule\per\meter}$. To facilitate the observation of coherent SW propagation over several microns, the Gilbert damping was artificially reduced to $\alpha = \num{1e-3}$.
A discretization of \SI{2}{\nano\meter} was used for one-dimensional simulations, and $\SI{4}{\nano\meter} \times \SI{4}{\nano\meter}$ for two-dimensional simulations to reduce computational cost. Periodic boundary conditions were applied in the $y$-direction in the one-dimensional case to emulate a semi-infinite film.

\paragraph{Simulation Setup.}
Simulations were conducted using Amumax~\cite{amumax2023}, our fork of MuMax3~\cite{mumax_2014}, solving the Landau–Lifshitz–Gilbert equation. SWs were excited by a uniform microwave field along $z$ with either a sinc-type profile (\SI{10}{\giga\hertz} cutoff) or a steady-state sine wave. Magnetization dynamics were sampled every \SI{16.66}{\pico\second} for \SI{100}{\nano\second} and processed via fast Fourier transform to extract spatial-frequency-resolved SW spectra. More details on the simulations and post-processing are available in the SI.

\section{Acknowledgment}
The research has received funding from the National Science Centre of Poland, Grant No. UMO-2020/37/B/ST3/03936 and 2023/49/N/ST3/03538, and the EU Research and Innovation Programme Horizon Europe (HORIZON-CL4-2021-DIGITAL-EMERGING-01) under grant agreement no. 101070347 (MANNGA). MM acknowledges funding from the Adam Mickiewicz University Foundation. 

\section{Associated content}
The data that support the findings of this study are available at: \url{https://doi.org/10.5281/zenodo.17068975}

\section{Conflict of Interest}
The authors declare no conflict of interest.

\bibliography{ref}



\section*{Supplementary material}

# Supporting Information: Efficient Generation of Second-Harmonic Propagating Spin Waves in a Thin, Out-of-Plane-Magnetized Ferromagnetic Film

Mathieu Moalic,* Youenn Patat, Mateusz Zelent, and Maciej Krawczyk

*Institute of Spintronics and Quantum Information, Faculty of Physics and Astronomy, Adam Mickiewicz University, 61-614 Poznań, Poland*

E-mail: matmoa@amu.edu.pl

## Methods

We are investigating a multilayered $[\text{Co/Pd}]_8$ thin film. In the 1D system, the length along the x-axis is 3400 nm, with a single unit cell along the y and z axes, and discretization of $2\times 2\times 13.2$ nm$^3$. We applied 1000 repetitions of periodic boundary conditions. The excitation region (ER) is located on the left side, with absorbing boundary conditions added to the right side for $x > 2600$ nm.

For the 2D case, we use a $1250 \times 1250 \times 1$ mesh, with discretization $4 \times 4 \times 13.2$ nm$^3$. The magnetic material is arranged in a disk with a radius of 2500 nm, and absorbing boundary conditions are applied for $\rho > 2000$ nm.

The multilayer film is modeled using an effective-medium approach, where the 8 repetitions of the Co/Pd bilayer stack (comprising Co of thickness 0.75 nm and Pd of thickness 0.9 nm) are represented as a single 13.2 nm-thick Co layer with effective material parameters,[1] as described in Ref.[2] We used the following parameters:[3] PMA constant $K_{u,\text{bulk}} = 4.5 \times 10^5$ J/m$^3$, saturation magnetization $M_\text{S} = 0.81\times 10^6$ A/m, exchange constant $A_\text{ex} = 1.3 \times 10^{-11}$ J/m. In most simulations, a low damping constant $\alpha = 1 \times 10^{-3}$ was used to produce a sharp spin-wave (SW) spectrum.

We discretized the system using $2 \times 2 \times 13.2$ nm$^3$ cuboids. To model the transition of the PMA value from the bulk value to 0, we employed a hyperbolic tangent function:

$$K_\text{u}(\rho) = \left(\frac{1}{2}\tanh\left(\frac{\rho - \rho_\text{edge}}{8}\right) + \frac{1}{2}\right) K_{u,\text{bulk}}$$

for $\rho > 100$ nm, where $\rho$ is the radial coordinate relative to the center of the antidot, and $\rho_\text{edge} = 150$ nm is the position of the rim edge, as shown in Fig.1(b). This function approximates the anisotropy-reduction profile typically found in experimental samples.

For micromagnetic simulations, we used our custom version of Mumax3,[4] called Amumax,[5] which solves the Landau-Lifshitz-Gilbert equation:

$$\frac{\text{d}\mathbf{m}}{\text{d}t} = \frac{\gamma\mu_0}{1 + \alpha^2}\left(\mathbf{m} \times \mathbf{H}_\text{eff} + \alpha\mathbf{m} \times (\mathbf{m} \times \mathbf{H}_\text{eff})\right), \quad (S1)$$

where $\mathbf{m} = \mathbf{M}/M_\text{S}$ is the normalized magnetization, $\mathbf{H}_\text{eff}$ is the effective magnetic field, $\gamma = 187 \times 10^9$ rad/(s·T) is the gyromagnetic ratio, and $\mu_0$ is the vacuum permeability. The effective magnetic field $\mathbf{H}_\text{eff}$ includes contributions from the demagnetizing field $\mathbf{H}_\text{d}$, the exchange field $\mathbf{H}_\text{exch}$, the uniaxial anisotropy field $\mathbf{H}_\text{anis}$, and the external magnetic field $\mathbf{H}_\text{ext}$. Thermal effects were neglected, and the effective field is expressed as:

$$\mathbf{H}_\text{eff} = \mathbf{H}_\text{d} + \mathbf{H}_\text{exch} + \mathbf{H}_\text{ext} + \mathbf{H}_\text{anis} + \mathbf{h}_\text{mf}, \quad (S2)$$

where the final term, $\mathbf{h}_{\mathrm{mf}}$, represents the microwave magnetic field used for SW excitation.

To obtain the SW spectrum, we collected space- and time-resolved in-plane magnetization data and applied a Hanning window along the time axis. We then performed a real discrete Fourier transform using the fast Fourier transform (FFT) algorithm along the time axis for each cuboid in the system.

## Spectral response inside the excitation region

Figure S1 (blue curve) shows the spin-wave spectrum under a sinusoidal drive at $f_0 = 8.60\,\mathrm{GHz}$. We set $B_0 = 354\,\mathrm{mT}$, $w = 70\,\mathrm{nm}$, and $b = 10\,\mathrm{mT}$, and computed the FFT of the $y$-component of the magnetization within the PR at $x = 2000\,\mathrm{nm}$. A dominant peak appears at $2f_0 = 17.20\,\mathrm{GHz}$ (magnitude 0.222 arb. units), while the fundamental at $f_0 = 8.60\,\mathrm{GHz}$ is much weaker (0.003 arb. units).

To test whether these discrete lines originate from a cavity-like response of the ER rather than domain-wall modes, we repeated the simulation (orange curve) and, after relaxation, pinned the spins for $x < 40\,\mathrm{nm}$. Suppressing the ER dynamics in this way removes the PR resonances.

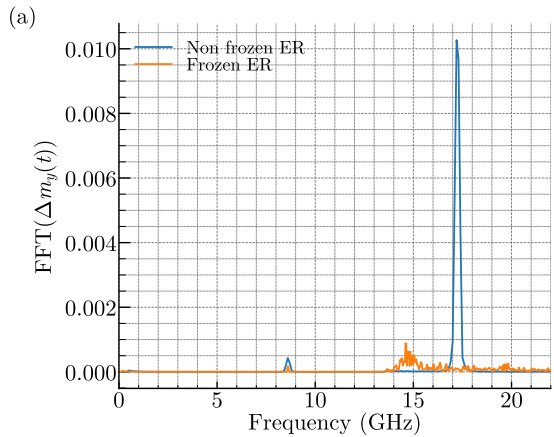

Figure S1: Spin-wave spectra measured in the PR at $x = 2000\,\mathrm{nm}$ following an out-of-plane sinc excitation ($f_{\mathrm{cut}} = 10\,\mathrm{GHz}$, peak field $10\,\mathrm{mT}$). Blue (unmodified system): ER fundamental at $f_0 = 8.60\,\mathrm{GHz}$, an additional ER-confined mode at $13.48\,\mathrm{GHz}$ (below the PR FMR threshold), and the second harmonic at $2f_0 = 17.20\,\mathrm{GHz}$; a weak, broad feature near $0.2\,\mathrm{GHz}$ arises from domain-wall–localized excitations. Orange (spins pinned for $x < 40\,\mathrm{nm}$): ER dynamics suppressed, leaving only low-frequency ($f < 2\,\mathrm{GHz}$) domain-wall modes.

## Short study on the effect of damping

We increased the damping parameter from the baseline value used throughout our results, $\alpha = 0.001$, to $\alpha = 0.01$. As shown in Fig. S2(a), this higher damping completely suppresses the generation of higher harmonics. When the pumping amplitude is increased from $6\,\mathrm{\mu T}$ to $60\,\mathrm{\mu T}$, as in Fig. S2(b), the behavior observed previously reappears: a clean $2f_0$ mode is generated, although it is progressively attenuated during propagation. Figure S2(c) demonstrates the presence of propagating spin waves for $b = 6\,\mathrm{\mu T}$ and $\alpha = 0.001$. For reference, Fig. S2(d) reproduces the simulation shown in Fig.2(b). These observations indicate the presence of a threshold effect, whereby the driving field amplitude must exceed a certain value in order to overcome damping.

13

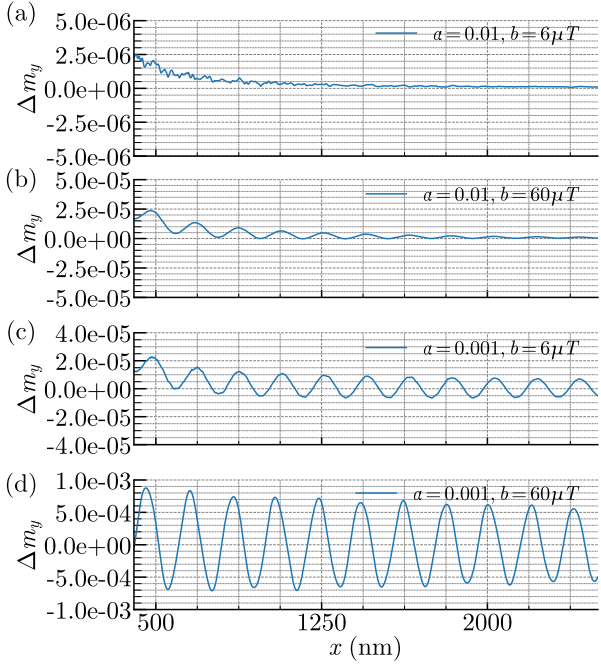

Figure S2: (a) Dynamical $y$-component of the magnetization along $x$ in the PR at steady state for an excitation field amplitude of $b = 6\,\mu\text{T}$ and a damping constant $\alpha = 0.01$. (b) Same as (a), but for $b = 60\,\mu\text{T}$ and $\alpha = 0.01$. (c) Same as (a), but for $b = 6\,\mu\text{T}$ and $\alpha = 0.001$. (d) Same as (a), but for $b = 60\,\mu\text{T}$ and $\alpha = 0.001$.

## High-frequency branch crossing

Figure S3(a) shows the PR response at $x = 2000\,\text{nm}$ following a *sinc* excitation along $z$ with a 10 GHz cutoff. We focus on the two branches that intersect at $w = 94\,\text{nm}$: the $2f_0$ branch and the ER eigensolution branch.

Figure S3(b) presents ER mode profiles for selected modes on the $2f_0$ branch. The number of ER nodal points increases from two to three at $w = 120\,\text{nm}$. In contrast, Fig. S3(c) shows that the ER eigensolution branch retains two ER nodal points as $w$ increases. At $w = 96\,\text{nm}$, the two mode profiles nearly coincide, enabling highly efficient $f_0$ to $2f_0$ conversion in the ER, as seen in Fig. 4(b).

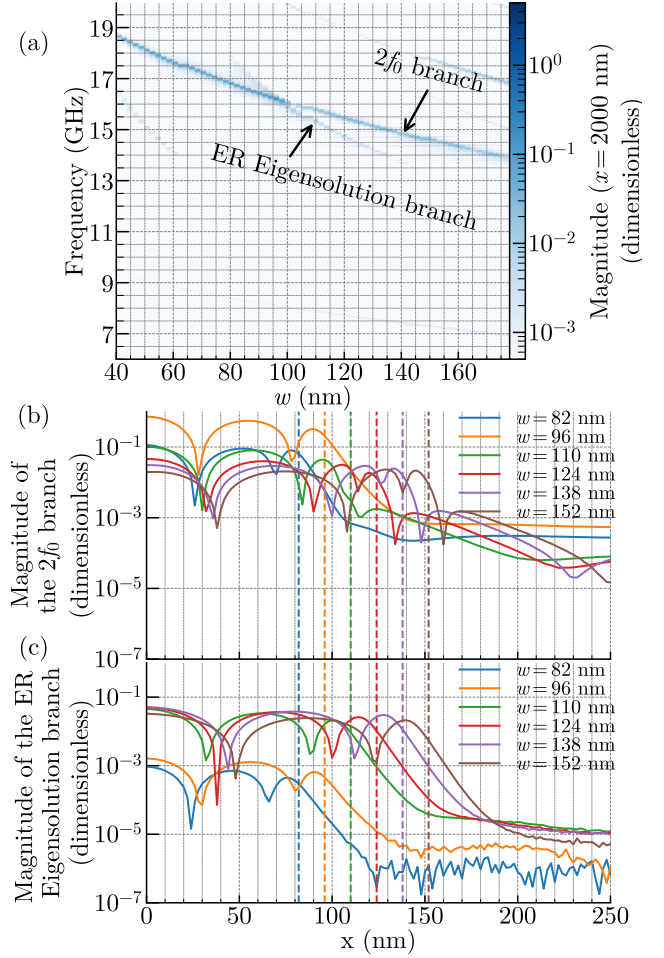

Figure S3: (a) Evolution of the spin-wave spectrum in the PR as the ER width $w$ is increased. (b) Spatial profiles of the normalized spin-wave amplitude corresponding to the $2f_0$ modes for given values of $w$. The dashed lines represent the corresponding $w$. (c) Same as (b), but for the ER Eigen-solution branch.


\end{document}